\documentclass[%
 reprint,
 amsmath,amssymb,
 aps,
]{revtex4-2}

\usepackage{comment}
\usepackage[normalem]{ulem}
\usepackage{graphicx}
\usepackage{dcolumn}
\usepackage{bm}
\usepackage{hyperref}

\usepackage{soul}
\usepackage[dvipsnames]{xcolor}
\setstcolor{red}

\begin{document}

\title{Universal singularities of anomalous diffusion in the Richardson class}

\author{Attilio L. Stella}
\affiliation{Department of Physics and Astronomy, University of Padova, Via Marzolo 8, I-35131 Padova, Italy}
\affiliation{INFN, Sezione di Padova, Via Marzolo 8, I-35131 Padova, Italy}
\author{Aleksei Chechkin}
\affiliation{Institute of Physics and Astronomy, University of Potsdam, D-14476 Potsdam-Golm, Germany,}
\affiliation{Faculty of Pure and Applied Mathematica, Hugo Steinhaus Center, University of Science and Technology, Wyspianskiego 27, 50-370 Wrocław, Poland,}
\affiliation{Akhiezer Institute for Theoretical Physics, 61108 Kharkov, Ukraine}
\author{Gianluca Teza}%
\email{gianluca.teza@weizmann.ac.il}
\affiliation{Department of Physics of Complex Systems, Weizmann Institute of Science, Rehovot 7610001, Israel}

\date{\today}

\begin{abstract}
Inhomogeneous environments are rather ubiquitous in nature, often implying anomalies resulting in deviation from Gaussianity of diffusion processes.
While sub- and superdiffusion are usually due to conversing environmental features (hindering or favoring the motion, respectively), they are both observed in systems ranging from the micro- to the cosmological scale.
Here we show how a model encompassing sub- and superdiffusion in an inhomogeneous environment exhibits a critical singularity in the normalized generator of the cumulants.
The singularity originates directly from the asymptotics of the non-Gaussian scaling function of displacement, which we prove to be independent of other details and hence to retain a universal character.
Our analysis, based on the method first applied in [A. L. Stella \emph{et al.}, \href{https://arxiv.org/abs/2209.02042}{arXiv:\textbf{2209.02042}} (2022)], further allows to establish a relation between the asympototics and diffusion exponents characteristic of processes in the Richardson class.
Extensive numerical tests fully confirm the results.
\end{abstract}

\maketitle

Anomalous spatial diffusion occurs when the mean squared displacement $\left< x^2 \right> \sim t^{2\nu}$ grows non-linearly in time, yielding by definition subdiffusion for $\nu< 1/2$ and superdiffusion when $\nu>1/2$ \cite{metzler2014anomalous}.
Deviations from normal diffusion ($\nu=1/2$) are often found in nature in systems ranging from microscopic to cosmological scales \cite{bouchaud1990anomalous}.
Subdiffusion ($\nu< 1/2$) is commonly observed in the biological contexts of particles moving inside living cells nuclei, cytoplasm and across membranes \cite{klages2008anomalous,metzler2000random,he2008random,sokolov2012models,metzler2014anomalous,golding2006physical,lubelski2008nonergodicity,weber2010bacterial,weigel2011ergodic,weber2012nonthermal,viswanathan1999optimizing}.
Superdiffusion ($\nu>1/2$) is also rather ubiquitous.
It is found in active intracellular transport \cite{goychuk2014molecular,caspi2000enhanced,caspi2000enhanced,arcizet2008temporal,duits2009mapping}, migration processes of cells \cite{dieterich2022anomalous} and more complex organisms and animals \cite{nathan2008movement,sims2008scaling,gonzalez2008understanding,viswanathan1996levy,viswanathan1999optimizing}, as well as in the contexts of target search processes \cite{barthelemy2008levy}, particle dispersion in turbulent fluids \cite{shlesinger1987levy,falkovich2001particles,boffetta2002relative}, and cosmic rays transport \cite{lagutin2003anomalous,uchaikin2013fractional}.

The probability density function (PDF) $p(x,t)$ of displacement $x$ is expected to satisfy at long times $t$
\begin{equation}\label{eq:scaling}
    p(x,t)\sim t^{-\nu} f(x/t^{\nu})
\end{equation}
where the scaling function $f(\cdot)$ has a non-Gaussian shape for $\nu\neq 1/2$ \cite{bouchaud1990anomalous}, implying an anomalous scaling of displacement in time \cite{cecconi2022probability}.
With $f(\cdot)$ integrable on $\mathbb{R}$ and decaying to zero sufficiently fast for large absolute argument, the $n$-th order cumulants of displacement diverge as $t^{n\nu}$ for $t\to\infty$.
Indeed, setting $z=x/t^{\nu}$, the form of asymptotic decay of the displacement scaling function can be argued to be \cite{stella2022anomalous}
\begin{equation}\label{eq:decay}
f(z) \sim |z|^\psi e^{-c |z|^{\delta+1}}
\end{equation}
for some positive constant $c$ and exponents $\delta$ and $\psi$.
Two known classes of anomalous diffusion processes, determined through specific  relations between the exponents $\delta$ and $\nu$, are expected to exhibit such a decay \cite{stella2022anomalous,cecconi2022probability}. The Fisher class
is characterized by the relation $\delta=\nu/(1-\nu)$, first established in the context of polymers with excluded volume in equilibrium \cite{fisher1966}, while the Richardson class relation, $\delta=(1-\nu)/\nu$, stems from a seminal paper dealing with particles dispersion in turbulent fluids \cite{richardson1926atmospheric}.
The latter is expected to apply when diffusion steps have certain dependencies on space position \cite{cherstvy2013}.

Anomalous scaling is also directly responsible for universal features of diffusion processes \cite{stella2022anomalous}.
The generating function $G(\lambda,t)=\int_{\mathbb{R}} dx\  e^{\lambda x} p(x,t)$ grows asymptotically as $\sim \exp (t^\zeta \varepsilon(\lambda))$ for some $\zeta>0$, defining a scaling cumulant generating function (SCGF)
\begin{equation}\label{eq:SCGF}
    \varepsilon(\lambda)=\lim_{t\to\infty}\frac{1}{t^{\zeta}} \log G(\lambda,t)
\end{equation}
which exhibits a power-law singularity around $\lambda=0$ depending on $\nu$ and $\delta$ \cite{stella2022anomalous}.
Universality is expected since the derivation shows that the singularity is determined by the asymptotic large $|z|$ behavior of the scaling function, which can be common to different processes.
Of such a feature, the model treated in the present work provides an explicit example.
The exponent $\zeta$ in Eq. \ref{eq:SCGF} determines the extensivity in time of the generating function.
The Fisher class is consistent with a standard definition of the SCGF, in which the generator is simply divided by $t$ (hence $\zeta=1$).
This extensivity in time reminds the extensivity in size one encounters when dealing with equilibrium critical phenomena, so that the $t\to\infty$ limit yields the analogue of a difference of equilibrium free energy densities, with time playing the role of size \cite{cardy2012,kadanoff2000statistical}.
For the Richardson class the method foresees a non-standard extensivity in time and the necessity to divide the generator by a power $t^\zeta$, with $\zeta\neq 1$ depending on the diffusion exponent \cite{stella2022anomalous}.
In spite of the different extensivity involved, also our derivation for Richardson processes should be regarded as a way of establishing a parallel between equilibrium criticality and dynamics \cite{stella2022anomalous}, according to a strategy on which much of our understanding of non-equilibrium is based \cite{Touchette2009,touchette2013,teza2022eigenvalue}.

The approach of Ref. \cite{stella2022anomalous} was explicitly applied and shown to predict exact results for the continuous time random walk (CTRW) model and fractional drift diffusion equations \cite{metzler2000random,montroll1965random,kenkre1973generalized}.
Both free and biased models exhibited sub-diffusion, while only in the biased case super-diffusion could be encompassed.
Moreover, such applications implied adoption of standard extensivity of the cumulant generator ($\zeta=1$ in Eq. \ref{eq:scgf_model}), as appropriate for processes in the Fisher class.
It remains an open issue to test the validity of this analysis for processes belonging to the Richardson class and possibly displaying both sub- and super-diffusion regimes.
The present work is devoted to the exploration of a specific diffusion model with both such features.

The process we consider in this work was introduced in Ref. \cite{cherstvy2013} to model a scenario of inhomonogenous diffusion, in which the diffusion constant has an explicit dependence on the position \cite{denisov2002statistical,denisov2002exactly}.
We show how this model can exhibit anomalous scaling at all times, implying that Eq. \ref{eq:scaling} holds as an equality.
However, unlike in the case of the CTRW model a direct analytical evaluation of the SCGF is not feasible for this process.
We show how the method of Ref. \cite{stella2022anomalous} allows to circumvent this problem and to correctly estimate the leading part of the SCGF, proven to abide by a non-trivial Richardson-like extensivity.
We highlight the existence of a universal singularity for the SCGF, as in the case of CTRW and fractional diffusion equations.
Integration with large deviation theory \cite{Touchette2009,touchette2013} shows how the PDF in the long-time limit is modulated by a non standard singular rate function, related to the fractional extensivity of the SCGF.
Ultimately, numerical evaluations of the integrals in the asymptotic regime corroborate the correctness of the predictions of our method.

The starting point is a particle moving in a one-dimensional axis according to the following Langevin dynamics:
\begin{equation}
    \frac{dx}{dt}=\sqrt{2D(x)}\xi(t)
\end{equation}
where $\xi$ is a $\delta$-correlated  ($\left<\xi(t)\xi(t')\right>=\delta(t-t')$)
white Gaussian noise, while the diffusion coefficient has a power-law spatial dependence $D(x)=D_0 |x|^{q}$ for some $D_0>0$ and any $q<2$.
Adopting Stratonovich prescription, the corresponding Fokker-Plank equation is:
\begin{equation}\label{eq:pdf_xt}
    \partial_t p(x,t) = \partial_x \left[\sqrt{D(x)}\partial_x \left[\sqrt{D(x)}p(x,t) \right] \right] 
\end{equation}
Given an initial condition $p(x,t=0)=\delta(x)$, the probability density function regulating the process can be shown to be \cite{cherstvy2013}
\begin{equation}
    p(x,t)=\frac{|x|^{-q/2}}{\sqrt{4\pi D_0t}}e^{-\frac{|x|^{2-q}}{(2-q)^2 D_0 t}}
\end{equation}
yielding a mean squared displacement
\begin{equation}
    \left<x^2(t) \right>=\frac{\Gamma \left( \frac{6-q}{2(2-q)} \right)}{\pi^{1/2}}
    (2-q)^{\frac{4}{2-q}}
    (D_0 t)^{\frac{2}{2-q}}
\end{equation}
where $\Gamma(\cdot)$ is the complete Gamma function.
It is therefore clear how this model provides subdiffusion in the case $q<0$ and superdiffusion for $0<q<2$, with the following relation connecting the spatial dependence of the diffusion constant with the diffusion exponent $\nu$:
\begin{equation}
    \nu=\frac{1}{2-q}\ .
\end{equation}
The PDF of the process can be easily seen to abide by the scaling form of Eq. \ref{eq:scaling} with
\begin{equation}\label{eq:scaling_function}
    f(z)=\frac{|z|^{\frac{1-2\nu}{\nu}}}{\sqrt{4\pi D_0}}e^{-\frac{\nu^2|z|^{1/\nu}}{ D_0 }}
\end{equation}
as scaling function, where we remind that $z=x/t^{\nu}$.
Note that scaling in Eq. \ref{eq:pdf_xt} holds exactly at all times, not just in the asymptotic long-time limit.
Quite remarkable is also the fact that the scaling function in Eq. \ref{eq:scaling_function} presents as valid on the whole $z$ axis the behavior predicted on the basis of general arguments in Ref. \cite{stella2022anomalous} for its large $|z|$ tails.
It can be shown that both these circumstances are determined by the particular initial condition chosen for the process \cite{sandev2022heterogeneous}.
Setting $p(x,0)=\delta(x-x_0)$ with some nonzero $x_0$ would lead to the validity of the scaling form in Eqs. \ref{eq:pdf_xt} and \ref{eq:scaling_function} only for large $t$ and large $|z|$ \cite{sandev2022heterogeneous,sandev2022stochastic}.

\begin{figure}
    \centering
    \includegraphics[width=0.99\linewidth]{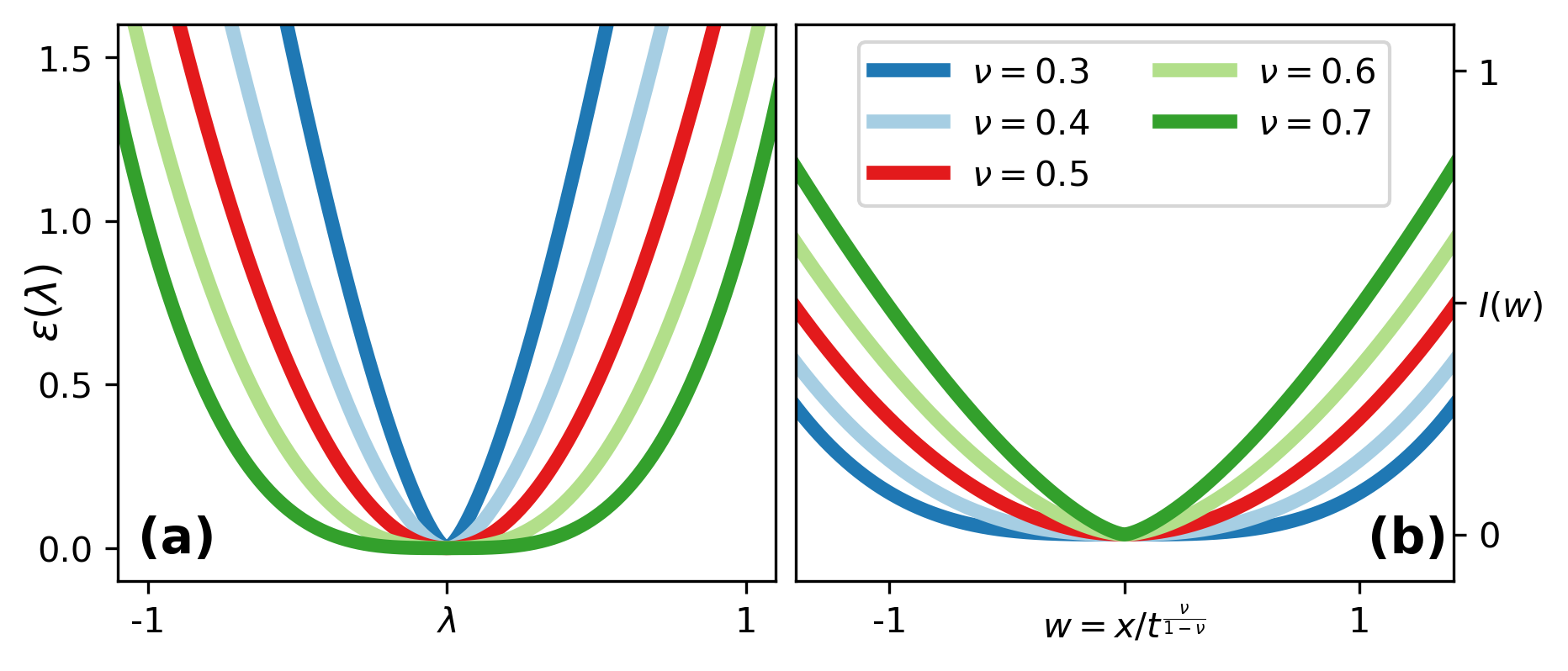}
    \caption{Examples of SCGFs $\varepsilon(\lambda)$ (a) and rate functions $I(w)$ (b) for different regimes of anomalous diffusion: subdiffusion (blue shades), superdiffusion (green shades) and normal diffusion (red).
    Both exhibit the expected power-law singularity predicted in Eqs. \ref{eq:scgf_model} and \ref{eq:rate_func_legendre} for $\lambda=0$ and $w=0$, respectively.}
    \label{fig:SCGF_rate}
\end{figure}

For every $0<\nu<1$ the generating function of the moments can be found through the two-sided Laplace transform $G(\lambda,t)=\int_{-\infty}^{+\infty} dx\ e^{\lambda x} p(x,t)$ \footnote{The case $\nu>1$ correspond to ``hyperballistic'' diffusion, in which the integral defining the generating function in Eq. \ref{eq:gen_func} diverges. We notice how, even though the Richardson class encompasses both regimes, the original Richardson model falls in the hyperballistic regime with $\nu=3/2$ \cite{richardson1926atmospheric,falkovich2001particles,boffetta2002relative}.}, which in terms of the rescaled displacement $z$ reads:
\begin{equation}\label{eq:gen_func}
    G(\lambda,t)= \frac{1}{\sqrt{4\pi D_0}} \int_{-\infty}^{+\infty} dz\ |z|^{\frac{1-2\nu}{\nu}} e^{\lambda z t^{\nu}-\frac{\nu^2|z|^{1/\nu}}{D_0}}
\end{equation}
An exact evaluation of this integral for long $t$ is not feasible, so that application of the Laplace's maximization method of Ref. \cite{stella2022anomalous} for its estimate, besides being suggested by the form of the tails, appears mandatory.

As time increases, the integrand in Eq. \ref{eq:gen_func} concentrates around some specific value $\bar{z}$ that maximizes the argument of the exponential. Separating the analysis for positive and negative values of $z$ we find
\begin{equation}\label{eq:zbar}
    \bar{z}=\textrm{sgn}(\lambda) \left( \frac{1}{\nu}D_0 |\lambda| t^{\nu} \right)^{\frac{\nu}{1-\nu}}
\end{equation}
where $\textrm{sgn}(\cdot)$ represents the sign function, implying that $\bar{z}$ and $\lambda$ have the same sign.
Moreover, for long times $\bar{z}$ diverges to $+\infty$ and $-\infty$ as a power of $t$ for $\lambda>0$ and $\lambda<0$, respectively.
Substituting such value in the exponential form and performing the Gaussian integration centered in $\bar{z}$ allows to obtain asymptotically \cite{stella2022anomalous}
\begin{eqnarray}\label{eq:logG_zbar}
    \log G(\lambda,t) &=\lambda t^{\nu} \bar{z}- \frac{\nu^2}{D_0} \bar{z}^{1/\nu}+ \\ \nonumber
    &+\frac{1}{2}\log(\frac{\nu}{1-\nu})
    +\mathcal{O}(\bar{z}^{- 1/\nu})
\end{eqnarray}
where a term proportional to $\log \bar{z}$ turns out to have coefficient zero. The cancellation of this term $\propto \log \bar{z}$ is due to the fact that, with reference to the notations adopted in Eq. \ref{eq:decay}, the exponents characterizing the tails of $f(z)$ satisfy $\psi =(\delta -1)/2$, which is also valid for all cases of anomalous diffusion studied in Ref. \cite{stella2022anomalous}.

Taking into account Eq. \ref{eq:zbar}, we can eventually write
\begin{eqnarray}\label{eq:logG_laplace}
    \log G(\lambda,t) 
    &= (1-\nu)\left( \frac{D_0}{\nu} t |\lambda|^{1/\nu} \right)^{\frac{\nu}{1-\nu}}+ \\ \nonumber &+\frac{1}{2}\log(\frac{\nu}{1-\nu})+\mathcal{O}(t^{-\frac{\nu}{1-\nu}})
\end{eqnarray}
implying a scaling of the cumulants of the Richardson class \cite{richardson1926atmospheric} with $\zeta=\nu/(1-\nu)$.
Consequently, a scaling cumulant generating function can be defined as
\begin{equation}\label{eq:scgf_model}
    \varepsilon(\lambda)=\lim_{t\to\infty} \frac{\log G(\lambda,t)}{t^{\frac{\nu}{1-\nu}}}=(1-\nu)\left( \frac{D_0}{\nu} |\lambda|^{1/\nu} \right)^{\frac{\nu}{1-\nu}}
\end{equation}
which exhibits a power-law singularity of order $1/(1-\nu)$ around $\lambda=0$ as shown above, implying a divergence of the $n$-th derivative as soon as $n$ exceeds $1/(1-\nu)$.
In the case $\nu=1/2$ the SCGF of the free Brownian diffusion is recovered, finding also consistency with the SCGF of a free Markovian (memory-less) CTRW \cite{teza2020exact,teza2020thesis}.

In Eq. \ref{eq:logG_laplace} appears a constant term $\frac{1}{2}\log \nu/(1-\nu)$ independent of time, which is negative for sub-, positive for super- and zero for normal diffusion.
In the context of equilibrium critical phenomena, this type of term, determined, e.g., by the specific form of the scaling function of the magnetization for finite magnetic systems at criticality, has been tentatively identified \cite{Bruce1995} with the Privman-Fisher \cite{Privman1984,blote1986} universal amplitude discussed in the context of finite size scaling theory \cite{cardy2012}.
In the present context time takes the place of size, but it appears remarkable that the term is nonzero only in case anomalous scaling holds ($\nu\neq 1/2$) and its sign marks a distinction between super- and subdiffusion. The parallel of the approach of Ref. \cite{stella2022anomalous} with studies of anomalous scaling in equilibrium critical phenomena certainly acquires motivation for deeper investigation in light of the presence of this analogue of Privman-Fisher amplitude.

\begin{figure}
    \centering
    \includegraphics[width=0.99\linewidth]{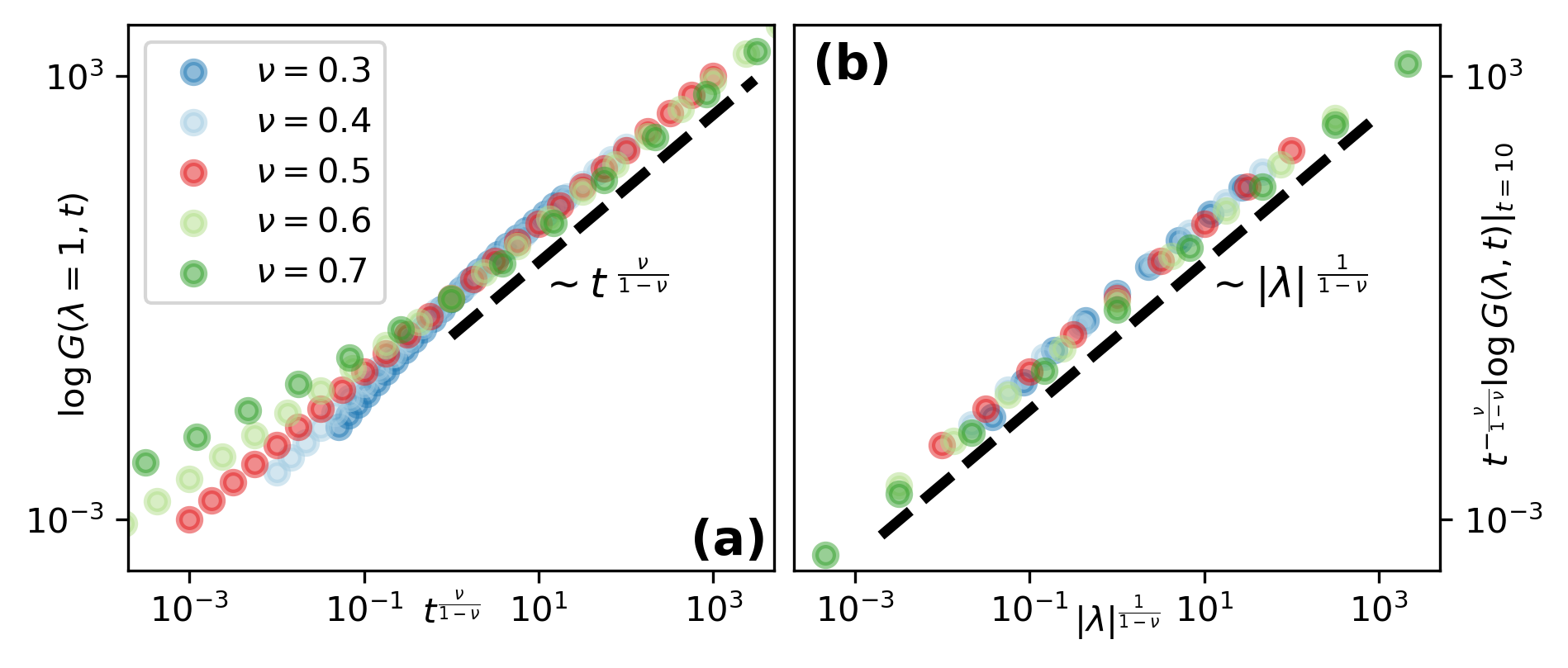}
    \caption{(a) Numerical evaluation of the cumulant generating function $\log G(\lambda=1,t)$ for different values of $\nu$ (including sub-, normal and superdiffusion).
    Plotting against the rescaled time $t^{\frac{\nu}{1-\nu}}$ shows an excellent collapse already at times $t>1$.
    (b) Numerical evaluation of the SCGF through the normalized cumulant generating function $t^{-\frac{\nu}{1-\nu}}\log G(\lambda,t)$ at $t=10$, hinting the presence of a Richardson kind of scaling for the cumulants.
    An excellent collapse for 6 decades hints that the SCGF $\varepsilon(\lambda)\sim |\lambda|^{1/(1-\nu)}$, implying a power-law singularity of such order around $\lambda=0$.
    }
    \label{fig:logG}
\end{figure}

Integration of our results within the framework of large deviation theory \cite{Touchette2009,touchette2013} shows how the singularity of the SCGF translates into a singularity of the rate function $I(w)$ modulating the probability of observing fluctuations of the rescaled position $w=x/t^{\frac{\nu}{1-\nu}}$ \cite{stella2022anomalous}.
In the case of normal diffusion ($\nu=1/2$), $w$ coincides dimensionally with a velocity, while for $\nu<1/2$ and $\nu>1/2$  can be interpreted as a sub- and super-velocity, respectively.
For simplicity, we will refer to $w$ as an ``anomalous velocity'' in this manuscript.
The probabiliy of observing a certain deviation from the typical value $w=0$ -- expected given the absence of any form of drift in the model -- in the long-time limit follows a large deviation principle
\begin{equation}
    p(x/t^{\frac{\nu}{1-\nu}}=w,t)\sim e^{-t^{\frac{\nu}{1-\nu}}I(w)}\ .
\end{equation}
The convexity and differentiability of the SCGF (Eq. \ref{eq:scgf_model}) ensures the validity of the G\"{a}rtner-Ellis theorem \cite{gartner1977on,ellis1984large} which allows to express the rate function as Legendre-Fenchel transform of $\varepsilon$ \cite{rockafellar1970convex,teza2020rate}:
\begin{equation}\label{eq:rate_func_legendre}
    I(w)=\sup_{\lambda\in\mathbb{R}}\left[ w\lambda-\varepsilon(\lambda) \right]=\frac{\nu^2|w|^{1/\nu}}{D_0}
\end{equation}
Thus, the anomalous scaling induces a singular behavior in the rate function, as already observed for processes in the Fisher class \cite{stella2022anomalous}.
It is convenient here to stress that the above result showing the consequences of anomalous scaling of the displacement distribution on the rate function is not related to what in the recent literature is referred to as ``anomalous scaling of dynamical large deviations''
\cite{nickelsen2018anomalous,smith2022anomalous,nickelsen2022noise}.

Finally, let us validate all the above results with numerical calculations.
Contrary to the CTRW and fractional drift diffusions examples presented in Ref. \cite{stella2022anomalous}, this inhomogeneous diffusion model does not allow for an exact evaluation of the cumulant generating function $\log G$.
The integral defining the generating function in Eq. \ref{eq:gen_func} cannot be expressed in terms of explicit functions for any arbitrary value of the diffusion exponent $0<\nu<1$.
Therefore, we need to proceed with a numerical estimation of such integral and extrapolate from the results its asymptotic dependence on time to verify that the extensivity of the cumulant generating function is the one predicted for the Richardson class.
In Fig. \ref{fig:logG}a we report the numerical evaluation of $\log G(\lambda=1,t)$ as a function of time, for different diffusion exponents ranging from $\nu=0.3$ (sub-diffusion) to $\nu=0.7$ (super-diffusion) including the case of normal diffusion $\nu=1/2$.
Plotting against $t^{\nu/(1-\nu)}$ in log-log scale, shows an excellent collapse on the bisector line already for $t\sim1$, quickly consolidating as time increases.
This corroborates the validity of the approach in estimating an extensivity of the Richardson class through the Laplace method (Eq. \ref{eq:logG_laplace}).
Additionally, the numerics find agreement with the predictions of the sign of the Privman-Fisher constant term appearing in Eq. \ref{eq:logG_laplace}.
For short times, we are able to appreciate how $\log G$ approaches the bisector line from below (negative constant) for sub-diffusive motions and from above (positive constant) for super-diffusive motions, while in the case of normal diffusion (zero-costant) the collapse holds at any time.

This result hints that for large enough times one should be able to normalize the cumulant generating function over $t^{\nu/(1-\nu)}$ and obtain a finite SCGF for all values of $\lambda$ (Eq. \ref{eq:scgf_model}).
We do so by evaluating numerically $\log G(\lambda,t)$ at $t=10$ as a function of the dual parameter $\lambda$, again for different values of $\nu$ encompassing sub-, normal and super-diffusion.
Normalizing such integral over $t^{\nu/(1-\nu)}$ as suggested by the previous analysis, we obtain an estimation of the SCGF, which is formally reached only in the $t\to\infty$ limit.
Plotting in log-log scale against $\lambda^{1/(1-\nu)}$ (Fig. \ref{fig:logG}b) we find a perfect collapse on the bisector line for all values of $\lambda$, simultaneously corroborating the full shape of the SCGF predicted in Eq. \ref{eq:scgf_model} and the existence of power-law singularities in the origin as those reported in Fig. \ref{fig:SCGF_rate}.

Summarizing, we showed that the method of Ref. \cite{stella2022anomalous} applies to a diffusion process in the Richardson class, predicting correctly the nonstandard extensivity in time of $\log G$ and the singularity of the SCGF in the dual parameter.
The model considered is remarkable in several respects.
In first place it satisfies scaling for all $t$ and presents the form in Eq. \ref{eq:scaling} of the scaling function on the whole $z$ axis.
The fact that these properties become only asymptotic for initial conditions different from $p(x,0)=\delta(x)$ provides a concrete example of the way universality mechanisms operate in the approach.
Indeed, the results of Ref. \cite{sandev2022heterogeneous} allow to easily verify that adoption of $p(x,0)=\delta(x-x_0)$ leaves scaling valid for $t\to\infty$ with the same form of scaling function at large $|z|$.
Thus, the leading singular behavior does not change for these modified initial conditions \cite{stella2022anomalous}.
Another remarkable feature of the model is the simple $\nu$-dependent form of the analogue of the Privman-Fisher amplitude, which distinguishes with its sign between sub- and super-diffusion. 
Once verified that the approach of Ref. \cite{stella2022anomalous} works successfully for processes in both the Fisher and the Richardson class, it is legitimate to ask if, in view of its flexibility, the range of applications could encompass also diffusions outside these classes.
The formalism leading to equations like Eq. \ref{eq:logG_zbar} in fact leaves room for different relations linking $\nu$ and $\delta$, only at the cost of adjusting the extensivity in time of $\log G$.
The exploration of such possibilities, or a deeper understanding of the reason why Fisher and Richardson relations play a special role is left for future investigations.

\begin{acknowledgments}
G. T. is supported by the Center for Statistical Mechanics at the Weizmann Institute of Science, the grant 662962 of the Simons foundation, the grants HALT and Hydrotronics of the EU Horizon 2020 program and the NSF-BSF grant 2020765. A. C. acknowledges support of the Polish National Agency for Academic Exchange (NAWA).
G.T. thanks Gregory Falkovich for useful discussions.
\end{acknowledgments}


\bibliography{refs.bib}

\providecommand{\noopsort}[1]{}\providecommand{\singleletter}[1]{#1}%
\begin{thebibliography}{56}%
\makeatletter
\providecommand \@ifxundefined [1]{%
 \@ifx{#1\undefined}
}%
\providecommand \@ifnum [1]{%
 \ifnum #1\expandafter \@firstoftwo
 \else \expandafter \@secondoftwo
 \fi
}%
\providecommand \@ifx [1]{%
 \ifx #1\expandafter \@firstoftwo
 \else \expandafter \@secondoftwo
 \fi
}%
\providecommand \natexlab [1]{#1}%
\providecommand \enquote  [1]{``#1''}%
\providecommand \bibnamefont  [1]{#1}%
\providecommand \bibfnamefont [1]{#1}%
\providecommand \citenamefont [1]{#1}%
\providecommand \href@noop [0]{\@secondoftwo}%
\providecommand \href [0]{\begingroup \@sanitize@url \@href}%
\providecommand \@href[1]{\@@startlink{#1}\@@href}%
\providecommand \@@href[1]{\endgroup#1\@@endlink}%
\providecommand \@sanitize@url [0]{\catcode `\\12\catcode `\$12\catcode
  `\&12\catcode `\#12\catcode `\^12\catcode `\_12\catcode `\%12\relax}%
\providecommand \@@startlink[1]{}%
\providecommand \@@endlink[0]{}%
\providecommand \url  [0]{\begingroup\@sanitize@url \@url }%
\providecommand \@url [1]{\endgroup\@href {#1}{\urlprefix }}%
\providecommand \urlprefix  [0]{URL }%
\providecommand \Eprint [0]{\href }%
\providecommand \doibase [0]{https://doi.org/}%
\providecommand \selectlanguage [0]{\@gobble}%
\providecommand \bibinfo  [0]{\@secondoftwo}%
\providecommand \bibfield  [0]{\@secondoftwo}%
\providecommand \translation [1]{[#1]}%
\providecommand \BibitemOpen [0]{}%
\providecommand \bibitemStop [0]{}%
\providecommand \bibitemNoStop [0]{.\EOS\space}%
\providecommand \EOS [0]{\spacefactor3000\relax}%
\providecommand \BibitemShut  [1]{\csname bibitem#1\endcsname}%
\let\auto@bib@innerbib\@empty
\bibitem [{\citenamefont {Metzler}\ \emph {et~al.}(2014)\citenamefont
  {Metzler}, \citenamefont {Jeon}, \citenamefont {Cherstvy},\ and\
  \citenamefont {Barkai}}]{metzler2014anomalous}%
  \BibitemOpen
  \bibfield  {author} {\bibinfo {author} {\bibfnamefont {R.}~\bibnamefont
  {Metzler}}, \bibinfo {author} {\bibfnamefont {J.-H.}\ \bibnamefont {Jeon}},
  \bibinfo {author} {\bibfnamefont {A.~G.}\ \bibnamefont {Cherstvy}},\ and\
  \bibinfo {author} {\bibfnamefont {E.}~\bibnamefont {Barkai}},\ }\bibfield
  {title} {\bibinfo {title} {Anomalous diffusion models and their properties:
  non-stationarity, non-ergodicity, and ageing at the centenary of single
  particle tracking},\ }\href {https://doi.org/10.1039/C4CP03465A} {\bibfield
  {journal} {\bibinfo  {journal} {Physical Chemistry Chemical Physics}\
  }\textbf {\bibinfo {volume} {16}},\ \bibinfo {pages} {24128} (\bibinfo {year}
  {2014})}\BibitemShut {NoStop}%
\bibitem [{\citenamefont {Bouchaud}\ and\ \citenamefont
  {Georges}(1990)}]{bouchaud1990anomalous}%
  \BibitemOpen
  \bibfield  {author} {\bibinfo {author} {\bibfnamefont {J.-P.}\ \bibnamefont
  {Bouchaud}}\ and\ \bibinfo {author} {\bibfnamefont {A.}~\bibnamefont
  {Georges}},\ }\bibfield  {title} {\bibinfo {title} {Anomalous diffusion in
  disordered media: statistical mechanisms, models and physical applications},\
  }\href {https://doi.org/10.1016/0370-1573(90)90099-N} {\bibfield  {journal}
  {\bibinfo  {journal} {Physics reports}\ }\textbf {\bibinfo {volume} {195}},\
  \bibinfo {pages} {127} (\bibinfo {year} {1990})}\BibitemShut {NoStop}%
\bibitem [{\citenamefont {Klages}\ \emph {et~al.}(2008)\citenamefont {Klages},
  \citenamefont {Radons},\ and\ \citenamefont {Sokolov}}]{klages2008anomalous}%
  \BibitemOpen
  \bibfield  {author} {\bibinfo {author} {\bibfnamefont {R.}~\bibnamefont
  {Klages}}, \bibinfo {author} {\bibfnamefont {G.}~\bibnamefont {Radons}},\
  and\ \bibinfo {author} {\bibfnamefont {I.~M.}\ \bibnamefont {Sokolov}},\
  }\href@noop {} {\emph {\bibinfo {title} {Anomalous transport}}}\ (\bibinfo
  {publisher} {Wiley Online Library},\ \bibinfo {year} {2008})\BibitemShut
  {NoStop}%
\bibitem [{\citenamefont {Metzler}\ and\ \citenamefont
  {Klafter}(2000)}]{metzler2000random}%
  \BibitemOpen
  \bibfield  {author} {\bibinfo {author} {\bibfnamefont {R.}~\bibnamefont
  {Metzler}}\ and\ \bibinfo {author} {\bibfnamefont {J.}~\bibnamefont
  {Klafter}},\ }\bibfield  {title} {\bibinfo {title} {The random walk's guide
  to anomalous diffusion: a fractional dynamics approach},\ }\href
  {https://doi.org/https://doi.org/10.1016/S0370-1573(00)00070-3} {\bibfield
  {journal} {\bibinfo  {journal} {Physics Reports}\ }\textbf {\bibinfo {volume}
  {339}},\ \bibinfo {pages} {1} (\bibinfo {year} {2000})}\BibitemShut {NoStop}%
\bibitem [{\citenamefont {He}\ \emph {et~al.}(2008)\citenamefont {He},
  \citenamefont {Burov}, \citenamefont {Metzler},\ and\ \citenamefont
  {Barkai}}]{he2008random}%
  \BibitemOpen
  \bibfield  {author} {\bibinfo {author} {\bibfnamefont {Y.}~\bibnamefont
  {He}}, \bibinfo {author} {\bibfnamefont {S.}~\bibnamefont {Burov}}, \bibinfo
  {author} {\bibfnamefont {R.}~\bibnamefont {Metzler}},\ and\ \bibinfo {author}
  {\bibfnamefont {E.}~\bibnamefont {Barkai}},\ }\bibfield  {title} {\bibinfo
  {title} {Random time-scale invariant diffusion and transport coefficients},\
  }\href {https://doi.org/10.1103/PhysRevLett.101.058101} {\bibfield  {journal}
  {\bibinfo  {journal} {Phys. Rev. Lett.}\ }\textbf {\bibinfo {volume} {101}},\
  \bibinfo {pages} {058101} (\bibinfo {year} {2008})}\BibitemShut {NoStop}%
\bibitem [{\citenamefont {Sokolov}(2012)}]{sokolov2012models}%
  \BibitemOpen
  \bibfield  {author} {\bibinfo {author} {\bibfnamefont {I.~M.}\ \bibnamefont
  {Sokolov}},\ }\bibfield  {title} {\bibinfo {title} {Models of anomalous
  diffusion in crowded environments},\ }\href
  {https://doi.org/10.1039/C2SM25701G} {\bibfield  {journal} {\bibinfo
  {journal} {Soft Matter}\ }\textbf {\bibinfo {volume} {8}},\ \bibinfo {pages}
  {9043} (\bibinfo {year} {2012})}\BibitemShut {NoStop}%
\bibitem [{\citenamefont {Golding}\ and\ \citenamefont
  {Cox}(2006)}]{golding2006physical}%
  \BibitemOpen
  \bibfield  {author} {\bibinfo {author} {\bibfnamefont {I.}~\bibnamefont
  {Golding}}\ and\ \bibinfo {author} {\bibfnamefont {E.~C.}\ \bibnamefont
  {Cox}},\ }\bibfield  {title} {\bibinfo {title} {Physical nature of bacterial
  cytoplasm},\ }\href {https://doi.org/10.1103/PhysRevLett.96.098102}
  {\bibfield  {journal} {\bibinfo  {journal} {Phys. Rev. Lett.}\ }\textbf
  {\bibinfo {volume} {96}},\ \bibinfo {pages} {098102} (\bibinfo {year}
  {2006})}\BibitemShut {NoStop}%
\bibitem [{\citenamefont {Lubelski}\ \emph {et~al.}(2008)\citenamefont
  {Lubelski}, \citenamefont {Sokolov},\ and\ \citenamefont
  {Klafter}}]{lubelski2008nonergodicity}%
  \BibitemOpen
  \bibfield  {author} {\bibinfo {author} {\bibfnamefont {A.}~\bibnamefont
  {Lubelski}}, \bibinfo {author} {\bibfnamefont {I.~M.}\ \bibnamefont
  {Sokolov}},\ and\ \bibinfo {author} {\bibfnamefont {J.}~\bibnamefont
  {Klafter}},\ }\bibfield  {title} {\bibinfo {title} {Nonergodicity mimics
  inhomogeneity in single particle tracking},\ }\href
  {https://doi.org/10.1103/PhysRevLett.100.250602} {\bibfield  {journal}
  {\bibinfo  {journal} {Phys. Rev. Lett.}\ }\textbf {\bibinfo {volume} {100}},\
  \bibinfo {pages} {250602} (\bibinfo {year} {2008})}\BibitemShut {NoStop}%
\bibitem [{\citenamefont {Weber}\ \emph {et~al.}(2010)\citenamefont {Weber},
  \citenamefont {Spakowitz},\ and\ \citenamefont
  {Theriot}}]{weber2010bacterial}%
  \BibitemOpen
  \bibfield  {author} {\bibinfo {author} {\bibfnamefont {S.~C.}\ \bibnamefont
  {Weber}}, \bibinfo {author} {\bibfnamefont {A.~J.}\ \bibnamefont
  {Spakowitz}},\ and\ \bibinfo {author} {\bibfnamefont {J.~A.}\ \bibnamefont
  {Theriot}},\ }\bibfield  {title} {\bibinfo {title} {Bacterial chromosomal
  loci move subdiffusively through a viscoelastic cytoplasm},\ }\href
  {https://doi.org/10.1103/PhysRevLett.104.238102} {\bibfield  {journal}
  {\bibinfo  {journal} {Phys. Rev. Lett.}\ }\textbf {\bibinfo {volume} {104}},\
  \bibinfo {pages} {238102} (\bibinfo {year} {2010})}\BibitemShut {NoStop}%
\bibitem [{\citenamefont {Weigel}\ \emph {et~al.}(2011)\citenamefont {Weigel},
  \citenamefont {Simon}, \citenamefont {Tamkun},\ and\ \citenamefont
  {Krapf}}]{weigel2011ergodic}%
  \BibitemOpen
  \bibfield  {author} {\bibinfo {author} {\bibfnamefont {A.~V.}\ \bibnamefont
  {Weigel}}, \bibinfo {author} {\bibfnamefont {B.}~\bibnamefont {Simon}},
  \bibinfo {author} {\bibfnamefont {M.~M.}\ \bibnamefont {Tamkun}},\ and\
  \bibinfo {author} {\bibfnamefont {D.}~\bibnamefont {Krapf}},\ }\bibfield
  {title} {\bibinfo {title} {Ergodic and nonergodic processes coexist in the
  plasma membrane as observed by single-molecule tracking},\ }\href
  {https://doi.org/10.1073/pnas.1016325108} {\bibfield  {journal} {\bibinfo
  {journal} {Proceedings of the National Academy of Sciences}\ }\textbf
  {\bibinfo {volume} {108}},\ \bibinfo {pages} {6438} (\bibinfo {year}
  {2011})}\BibitemShut {NoStop}%
\bibitem [{\citenamefont {Weber}\ \emph {et~al.}(2012)\citenamefont {Weber},
  \citenamefont {Spakowitz},\ and\ \citenamefont
  {Theriot}}]{weber2012nonthermal}%
  \BibitemOpen
  \bibfield  {author} {\bibinfo {author} {\bibfnamefont {S.~C.}\ \bibnamefont
  {Weber}}, \bibinfo {author} {\bibfnamefont {A.~J.}\ \bibnamefont
  {Spakowitz}},\ and\ \bibinfo {author} {\bibfnamefont {J.~A.}\ \bibnamefont
  {Theriot}},\ }\bibfield  {title} {\bibinfo {title} {Nonthermal atp-dependent
  fluctuations contribute to the in vivo motion of chromosomal loci},\ }\href
  {https://doi.org/10.1073/pnas.1119505109} {\bibfield  {journal} {\bibinfo
  {journal} {Proceedings of the National Academy of Sciences}\ }\textbf
  {\bibinfo {volume} {109}},\ \bibinfo {pages} {7338} (\bibinfo {year}
  {2012})}\BibitemShut {NoStop}%
\bibitem [{\citenamefont {Viswanathan}\ \emph {et~al.}(1999)\citenamefont
  {Viswanathan}, \citenamefont {Buldyrev}, \citenamefont {Havlin},
  \citenamefont {Da~Luz}, \citenamefont {Raposo},\ and\ \citenamefont
  {Stanley}}]{viswanathan1999optimizing}%
  \BibitemOpen
  \bibfield  {author} {\bibinfo {author} {\bibfnamefont {G.~M.}\ \bibnamefont
  {Viswanathan}}, \bibinfo {author} {\bibfnamefont {S.~V.}\ \bibnamefont
  {Buldyrev}}, \bibinfo {author} {\bibfnamefont {S.}~\bibnamefont {Havlin}},
  \bibinfo {author} {\bibfnamefont {M.}~\bibnamefont {Da~Luz}}, \bibinfo
  {author} {\bibfnamefont {E.}~\bibnamefont {Raposo}},\ and\ \bibinfo {author}
  {\bibfnamefont {H.~E.}\ \bibnamefont {Stanley}},\ }\bibfield  {title}
  {\bibinfo {title} {Optimizing the success of random searches},\ }\href
  {https://doi.org/10.1038/44831} {\bibfield  {journal} {\bibinfo  {journal}
  {nature}\ }\textbf {\bibinfo {volume} {401}},\ \bibinfo {pages} {911}
  (\bibinfo {year} {1999})}\BibitemShut {NoStop}%
\bibitem [{\citenamefont {Goychuk}\ \emph {et~al.}(2014)\citenamefont
  {Goychuk}, \citenamefont {Kharchenko},\ and\ \citenamefont
  {Metzler}}]{goychuk2014molecular}%
  \BibitemOpen
  \bibfield  {author} {\bibinfo {author} {\bibfnamefont {I.}~\bibnamefont
  {Goychuk}}, \bibinfo {author} {\bibfnamefont {V.~O.}\ \bibnamefont
  {Kharchenko}},\ and\ \bibinfo {author} {\bibfnamefont {R.}~\bibnamefont
  {Metzler}},\ }\bibfield  {title} {\bibinfo {title} {Molecular motors pulling
  cargos in the viscoelastic cytosol: how power strokes beat subdiffusion},\
  }\href {https://doi.org/10.1039/C4CP01234H} {\bibfield  {journal} {\bibinfo
  {journal} {Physical Chemistry Chemical Physics}\ }\textbf {\bibinfo {volume}
  {16}},\ \bibinfo {pages} {16524} (\bibinfo {year} {2014})}\BibitemShut
  {NoStop}%
\bibitem [{\citenamefont {Caspi}\ \emph {et~al.}(2000)\citenamefont {Caspi},
  \citenamefont {Granek},\ and\ \citenamefont {Elbaum}}]{caspi2000enhanced}%
  \BibitemOpen
  \bibfield  {author} {\bibinfo {author} {\bibfnamefont {A.}~\bibnamefont
  {Caspi}}, \bibinfo {author} {\bibfnamefont {R.}~\bibnamefont {Granek}},\ and\
  \bibinfo {author} {\bibfnamefont {M.}~\bibnamefont {Elbaum}},\ }\bibfield
  {title} {\bibinfo {title} {Enhanced diffusion in active intracellular
  transport},\ }\href {https://doi.org/10.1103/PhysRevLett.85.5655} {\bibfield
  {journal} {\bibinfo  {journal} {Phys. Rev. Lett.}\ }\textbf {\bibinfo
  {volume} {85}},\ \bibinfo {pages} {5655} (\bibinfo {year}
  {2000})}\BibitemShut {NoStop}%
\bibitem [{\citenamefont {Arcizet}\ \emph {et~al.}(2008)\citenamefont
  {Arcizet}, \citenamefont {Meier}, \citenamefont {Sackmann}, \citenamefont
  {R\"adler},\ and\ \citenamefont {Heinrich}}]{arcizet2008temporal}%
  \BibitemOpen
  \bibfield  {author} {\bibinfo {author} {\bibfnamefont {D.}~\bibnamefont
  {Arcizet}}, \bibinfo {author} {\bibfnamefont {B.}~\bibnamefont {Meier}},
  \bibinfo {author} {\bibfnamefont {E.}~\bibnamefont {Sackmann}}, \bibinfo
  {author} {\bibfnamefont {J.~O.}\ \bibnamefont {R\"adler}},\ and\ \bibinfo
  {author} {\bibfnamefont {D.}~\bibnamefont {Heinrich}},\ }\bibfield  {title}
  {\bibinfo {title} {Temporal analysis of active and passive transport in
  living cells},\ }\href {https://doi.org/10.1103/PhysRevLett.101.248103}
  {\bibfield  {journal} {\bibinfo  {journal} {Phys. Rev. Lett.}\ }\textbf
  {\bibinfo {volume} {101}},\ \bibinfo {pages} {248103} (\bibinfo {year}
  {2008})}\BibitemShut {NoStop}%
\bibitem [{\citenamefont {Duits}\ \emph {et~al.}(2009)\citenamefont {Duits},
  \citenamefont {Li}, \citenamefont {Vanapalli},\ and\ \citenamefont
  {Mugele}}]{duits2009mapping}%
  \BibitemOpen
  \bibfield  {author} {\bibinfo {author} {\bibfnamefont {M.~H.~G.}\
  \bibnamefont {Duits}}, \bibinfo {author} {\bibfnamefont {Y.}~\bibnamefont
  {Li}}, \bibinfo {author} {\bibfnamefont {S.~A.}\ \bibnamefont {Vanapalli}},\
  and\ \bibinfo {author} {\bibfnamefont {F.}~\bibnamefont {Mugele}},\
  }\bibfield  {title} {\bibinfo {title} {Mapping of spatiotemporal
  heterogeneous particle dynamics in living cells},\ }\href
  {https://doi.org/10.1103/PhysRevE.79.051910} {\bibfield  {journal} {\bibinfo
  {journal} {Phys. Rev. E}\ }\textbf {\bibinfo {volume} {79}},\ \bibinfo
  {pages} {051910} (\bibinfo {year} {2009})}\BibitemShut {NoStop}%
\bibitem [{\citenamefont {Dieterich}\ \emph {et~al.}(2022)\citenamefont
  {Dieterich}, \citenamefont {Lindemann}, \citenamefont {Moskopp},
  \citenamefont {Tauzin}, \citenamefont {Huttenlocher}, \citenamefont {Klages},
  \citenamefont {Chechkin},\ and\ \citenamefont
  {Schwab}}]{dieterich2022anomalous}%
  \BibitemOpen
  \bibfield  {author} {\bibinfo {author} {\bibfnamefont {P.}~\bibnamefont
  {Dieterich}}, \bibinfo {author} {\bibfnamefont {O.}~\bibnamefont
  {Lindemann}}, \bibinfo {author} {\bibfnamefont {M.~L.}\ \bibnamefont
  {Moskopp}}, \bibinfo {author} {\bibfnamefont {S.}~\bibnamefont {Tauzin}},
  \bibinfo {author} {\bibfnamefont {A.}~\bibnamefont {Huttenlocher}}, \bibinfo
  {author} {\bibfnamefont {R.}~\bibnamefont {Klages}}, \bibinfo {author}
  {\bibfnamefont {A.}~\bibnamefont {Chechkin}},\ and\ \bibinfo {author}
  {\bibfnamefont {A.}~\bibnamefont {Schwab}},\ }\bibfield  {title} {\bibinfo
  {title} {Anomalous diffusion and asymmetric tempering memory in neutrophil
  chemotaxis},\ }\href {https://doi.org/10.1371/journal.pcbi.1010089}
  {\bibfield  {journal} {\bibinfo  {journal} {PLOS Computational Biology}\
  }\textbf {\bibinfo {volume} {18}},\ \bibinfo {pages} {e1010089} (\bibinfo
  {year} {2022})}\BibitemShut {NoStop}%
\bibitem [{\citenamefont {Nathan}\ \emph {et~al.}(2008)\citenamefont {Nathan},
  \citenamefont {Getz}, \citenamefont {Revilla}, \citenamefont {Holyoak},
  \citenamefont {Kadmon}, \citenamefont {Saltz},\ and\ \citenamefont
  {Smouse}}]{nathan2008movement}%
  \BibitemOpen
  \bibfield  {author} {\bibinfo {author} {\bibfnamefont {R.}~\bibnamefont
  {Nathan}}, \bibinfo {author} {\bibfnamefont {W.~M.}\ \bibnamefont {Getz}},
  \bibinfo {author} {\bibfnamefont {E.}~\bibnamefont {Revilla}}, \bibinfo
  {author} {\bibfnamefont {M.}~\bibnamefont {Holyoak}}, \bibinfo {author}
  {\bibfnamefont {R.}~\bibnamefont {Kadmon}}, \bibinfo {author} {\bibfnamefont
  {D.}~\bibnamefont {Saltz}},\ and\ \bibinfo {author} {\bibfnamefont {P.~E.}\
  \bibnamefont {Smouse}},\ }\bibfield  {title} {\bibinfo {title} {A movement
  ecology paradigm for unifying organismal movement research},\ }\href
  {https://doi.org/10.1073/pnas.080037510} {\bibfield  {journal} {\bibinfo
  {journal} {Proceedings of the National Academy of Sciences}\ }\textbf
  {\bibinfo {volume} {105}},\ \bibinfo {pages} {19052} (\bibinfo {year}
  {2008})}\BibitemShut {NoStop}%
\bibitem [{\citenamefont {Sims}\ \emph {et~al.}(2008)\citenamefont {Sims},
  \citenamefont {Southall}, \citenamefont {Humphries}, \citenamefont {Hays},
  \citenamefont {Bradshaw}, \citenamefont {Pitchford}, \citenamefont {James},
  \citenamefont {Ahmed}, \citenamefont {Brierley}, \citenamefont {Hindell}
  \emph {et~al.}}]{sims2008scaling}%
  \BibitemOpen
  \bibfield  {author} {\bibinfo {author} {\bibfnamefont {D.~W.}\ \bibnamefont
  {Sims}}, \bibinfo {author} {\bibfnamefont {E.~J.}\ \bibnamefont {Southall}},
  \bibinfo {author} {\bibfnamefont {N.~E.}\ \bibnamefont {Humphries}}, \bibinfo
  {author} {\bibfnamefont {G.~C.}\ \bibnamefont {Hays}}, \bibinfo {author}
  {\bibfnamefont {C.~J.}\ \bibnamefont {Bradshaw}}, \bibinfo {author}
  {\bibfnamefont {J.~W.}\ \bibnamefont {Pitchford}}, \bibinfo {author}
  {\bibfnamefont {A.}~\bibnamefont {James}}, \bibinfo {author} {\bibfnamefont
  {M.~Z.}\ \bibnamefont {Ahmed}}, \bibinfo {author} {\bibfnamefont {A.~S.}\
  \bibnamefont {Brierley}}, \bibinfo {author} {\bibfnamefont {M.~A.}\
  \bibnamefont {Hindell}}, \emph {et~al.},\ }\bibfield  {title} {\bibinfo
  {title} {Scaling laws of marine predator search behaviour},\ }\href
  {https://doi.org/10.1038/nature06518} {\bibfield  {journal} {\bibinfo
  {journal} {Nature}\ }\textbf {\bibinfo {volume} {451}},\ \bibinfo {pages}
  {1098} (\bibinfo {year} {2008})}\BibitemShut {NoStop}%
\bibitem [{\citenamefont {Gonzalez}\ \emph {et~al.}(2008)\citenamefont
  {Gonzalez}, \citenamefont {Hidalgo},\ and\ \citenamefont
  {Barabasi}}]{gonzalez2008understanding}%
  \BibitemOpen
  \bibfield  {author} {\bibinfo {author} {\bibfnamefont {M.~C.}\ \bibnamefont
  {Gonzalez}}, \bibinfo {author} {\bibfnamefont {C.~A.}\ \bibnamefont
  {Hidalgo}},\ and\ \bibinfo {author} {\bibfnamefont {A.-L.}\ \bibnamefont
  {Barabasi}},\ }\bibfield  {title} {\bibinfo {title} {Understanding individual
  human mobility patterns},\ }\href {https://doi.org/10.1038/nature06958}
  {\bibfield  {journal} {\bibinfo  {journal} {nature}\ }\textbf {\bibinfo
  {volume} {453}},\ \bibinfo {pages} {779} (\bibinfo {year}
  {2008})}\BibitemShut {NoStop}%
\bibitem [{\citenamefont {Viswanathan}\ \emph {et~al.}(1996)\citenamefont
  {Viswanathan}, \citenamefont {Afanasyev}, \citenamefont {Buldyrev},
  \citenamefont {Murphy}, \citenamefont {Prince},\ and\ \citenamefont
  {Stanley}}]{viswanathan1996levy}%
  \BibitemOpen
  \bibfield  {author} {\bibinfo {author} {\bibfnamefont {G.~M.}\ \bibnamefont
  {Viswanathan}}, \bibinfo {author} {\bibfnamefont {V.}~\bibnamefont
  {Afanasyev}}, \bibinfo {author} {\bibfnamefont {S.~V.}\ \bibnamefont
  {Buldyrev}}, \bibinfo {author} {\bibfnamefont {E.~J.}\ \bibnamefont
  {Murphy}}, \bibinfo {author} {\bibfnamefont {P.~A.}\ \bibnamefont {Prince}},\
  and\ \bibinfo {author} {\bibfnamefont {H.~E.}\ \bibnamefont {Stanley}},\
  }\bibfield  {title} {\bibinfo {title} {L{\'e}vy flight search patterns of
  wandering albatrosses},\ }\href {https://doi.org/10.1038/381413a0} {\bibfield
   {journal} {\bibinfo  {journal} {Nature}\ }\textbf {\bibinfo {volume}
  {381}},\ \bibinfo {pages} {413} (\bibinfo {year} {1996})}\BibitemShut
  {NoStop}%
\bibitem [{\citenamefont {Barthelemy}\ \emph {et~al.}(2008)\citenamefont
  {Barthelemy}, \citenamefont {Bertolotti},\ and\ \citenamefont
  {Wiersma}}]{barthelemy2008levy}%
  \BibitemOpen
  \bibfield  {author} {\bibinfo {author} {\bibfnamefont {P.}~\bibnamefont
  {Barthelemy}}, \bibinfo {author} {\bibfnamefont {J.}~\bibnamefont
  {Bertolotti}},\ and\ \bibinfo {author} {\bibfnamefont {D.~S.}\ \bibnamefont
  {Wiersma}},\ }\bibfield  {title} {\bibinfo {title} {A l{\'e}vy flight for
  light},\ }\href {https://doi.org/10.1038/nature06948} {\bibfield  {journal}
  {\bibinfo  {journal} {Nature}\ }\textbf {\bibinfo {volume} {453}},\ \bibinfo
  {pages} {495} (\bibinfo {year} {2008})}\BibitemShut {NoStop}%
\bibitem [{\citenamefont {Shlesinger}\ \emph {et~al.}(1987)\citenamefont
  {Shlesinger}, \citenamefont {West},\ and\ \citenamefont
  {Klafter}}]{shlesinger1987levy}%
  \BibitemOpen
  \bibfield  {author} {\bibinfo {author} {\bibfnamefont {M.~F.}\ \bibnamefont
  {Shlesinger}}, \bibinfo {author} {\bibfnamefont {B.~J.}\ \bibnamefont
  {West}},\ and\ \bibinfo {author} {\bibfnamefont {J.}~\bibnamefont
  {Klafter}},\ }\bibfield  {title} {\bibinfo {title} {L\'evy dynamics of
  enhanced diffusion: Application to turbulence},\ }\href
  {https://doi.org/10.1103/PhysRevLett.58.1100} {\bibfield  {journal} {\bibinfo
   {journal} {Phys. Rev. Lett.}\ }\textbf {\bibinfo {volume} {58}},\ \bibinfo
  {pages} {1100} (\bibinfo {year} {1987})}\BibitemShut {NoStop}%
\bibitem [{\citenamefont {Falkovich}\ \emph {et~al.}(2001)\citenamefont
  {Falkovich}, \citenamefont {Gaw\ifmmode~\mbox{\c{e}}\else \c{e}\fi{}dzki},\
  and\ \citenamefont {Vergassola}}]{falkovich2001particles}%
  \BibitemOpen
  \bibfield  {author} {\bibinfo {author} {\bibfnamefont {G.}~\bibnamefont
  {Falkovich}}, \bibinfo {author} {\bibfnamefont {K.}~\bibnamefont
  {Gaw\ifmmode~\mbox{\c{e}}\else \c{e}\fi{}dzki}},\ and\ \bibinfo {author}
  {\bibfnamefont {M.}~\bibnamefont {Vergassola}},\ }\bibfield  {title}
  {\bibinfo {title} {Particles and fields in fluid turbulence},\ }\href
  {https://doi.org/10.1103/RevModPhys.73.913} {\bibfield  {journal} {\bibinfo
  {journal} {Rev. Mod. Phys.}\ }\textbf {\bibinfo {volume} {73}},\ \bibinfo
  {pages} {913} (\bibinfo {year} {2001})}\BibitemShut {NoStop}%
\bibitem [{\citenamefont {Boffetta}\ and\ \citenamefont
  {Sokolov}(2002)}]{boffetta2002relative}%
  \BibitemOpen
  \bibfield  {author} {\bibinfo {author} {\bibfnamefont {G.}~\bibnamefont
  {Boffetta}}\ and\ \bibinfo {author} {\bibfnamefont {I.~M.}\ \bibnamefont
  {Sokolov}},\ }\bibfield  {title} {\bibinfo {title} {Relative dispersion in
  fully developed turbulence: The richardson's law and intermittency
  corrections},\ }\href {https://doi.org/10.1103/PhysRevLett.88.094501}
  {\bibfield  {journal} {\bibinfo  {journal} {Phys. Rev. Lett.}\ }\textbf
  {\bibinfo {volume} {88}},\ \bibinfo {pages} {094501} (\bibinfo {year}
  {2002})}\BibitemShut {NoStop}%
\bibitem [{\citenamefont {Lagutin}\ and\ \citenamefont
  {Uchaikin}(2003)}]{lagutin2003anomalous}%
  \BibitemOpen
  \bibfield  {author} {\bibinfo {author} {\bibfnamefont {A.}~\bibnamefont
  {Lagutin}}\ and\ \bibinfo {author} {\bibfnamefont {V.}~\bibnamefont
  {Uchaikin}},\ }\bibfield  {title} {\bibinfo {title} {Anomalous diffusion
  equation: Application to cosmic ray transport},\ }\href
  {https://doi.org/10.1016/S0168-583X(02)01362-9} {\bibfield  {journal}
  {\bibinfo  {journal} {Nuclear Instruments and Methods in Physics Research
  Section B: Beam Interactions with Materials and Atoms}\ }\textbf {\bibinfo
  {volume} {201}},\ \bibinfo {pages} {212} (\bibinfo {year}
  {2003})}\BibitemShut {NoStop}%
\bibitem [{\citenamefont {Uchaikin}(2013)}]{uchaikin2013fractional}%
  \BibitemOpen
  \bibfield  {author} {\bibinfo {author} {\bibfnamefont {V.~V.}\ \bibnamefont
  {Uchaikin}},\ }\bibfield  {title} {\bibinfo {title} {Fractional phenomenology
  of cosmic ray anomalous diffusion},\ }\href
  {https://doi.org/10.3367/ufne.0183.201311b.1175} {\bibfield  {journal}
  {\bibinfo  {journal} {Physics-Uspekhi}\ }\textbf {\bibinfo {volume} {56}},\
  \bibinfo {pages} {1074} (\bibinfo {year} {2013})}\BibitemShut {NoStop}%
\bibitem [{\citenamefont {Cecconi}\ \emph {et~al.}(2022)\citenamefont
  {Cecconi}, \citenamefont {Costantini}, \citenamefont {Taloni},\ and\
  \citenamefont {Vulpiani}}]{cecconi2022probability}%
  \BibitemOpen
  \bibfield  {author} {\bibinfo {author} {\bibfnamefont {F.}~\bibnamefont
  {Cecconi}}, \bibinfo {author} {\bibfnamefont {G.}~\bibnamefont {Costantini}},
  \bibinfo {author} {\bibfnamefont {A.}~\bibnamefont {Taloni}},\ and\ \bibinfo
  {author} {\bibfnamefont {A.}~\bibnamefont {Vulpiani}},\ }\bibfield  {title}
  {\bibinfo {title} {Probability distribution functions of sub- and
  superdiffusive systems},\ }\href
  {https://doi.org/10.1103/PhysRevResearch.4.023192} {\bibfield  {journal}
  {\bibinfo  {journal} {Phys. Rev. Research}\ }\textbf {\bibinfo {volume}
  {4}},\ \bibinfo {pages} {023192} (\bibinfo {year} {2022})}\BibitemShut
  {NoStop}%
\bibitem [{\citenamefont {Stella}\ \emph {et~al.}(2022)\citenamefont {Stella},
  \citenamefont {Chechkin},\ and\ \citenamefont {Teza}}]{stella2022anomalous}%
  \BibitemOpen
  \bibfield  {author} {\bibinfo {author} {\bibfnamefont {A.~L.}\ \bibnamefont
  {Stella}}, \bibinfo {author} {\bibfnamefont {A.}~\bibnamefont {Chechkin}},\
  and\ \bibinfo {author} {\bibfnamefont {G.}~\bibnamefont {Teza}},\ }\bibfield
  {title} {\bibinfo {title} {Anomalous dynamical scaling determines universal
  critical singularities},\ }\bibfield  {journal} {\bibinfo  {journal} {arXiv
  preprint arXiv:2209.02042}\ }\href
  {https://doi.org/10.48550/arXiv.2209.02042} {10.48550/arXiv.2209.02042}
  (\bibinfo {year} {2022})\BibitemShut {NoStop}%
\bibitem [{\citenamefont {Fisher}(1966)}]{fisher1966}%
  \BibitemOpen
  \bibfield  {author} {\bibinfo {author} {\bibfnamefont {M.~E.}\ \bibnamefont
  {Fisher}},\ }\bibfield  {title} {\bibinfo {title} {Shape of a self‐avoiding
  walk or polymer chain},\ }\href {https://doi.org/10.1063/1.1726734}
  {\bibfield  {journal} {\bibinfo  {journal} {The Journal of Chemical Physics}\
  }\textbf {\bibinfo {volume} {44}},\ \bibinfo {pages} {616} (\bibinfo {year}
  {1966})}\BibitemShut {NoStop}%
\bibitem [{\citenamefont {Richardson}\ and\ \citenamefont
  {Walker}(1926)}]{richardson1926atmospheric}%
  \BibitemOpen
  \bibfield  {author} {\bibinfo {author} {\bibfnamefont {L.~F.}\ \bibnamefont
  {Richardson}}\ and\ \bibinfo {author} {\bibfnamefont {G.~T.}\ \bibnamefont
  {Walker}},\ }\bibfield  {title} {\bibinfo {title} {Atmospheric diffusion
  shown on a distance-neighbour graph},\ }\href
  {https://doi.org/10.1098/rspa.1926.0043} {\bibfield  {journal} {\bibinfo
  {journal} {Proceedings of the Royal Society of London. Series A, Containing
  Papers of a Mathematical and Physical Character}\ }\textbf {\bibinfo {volume}
  {110}},\ \bibinfo {pages} {709} (\bibinfo {year} {1926})}\BibitemShut
  {NoStop}%
\bibitem [{\citenamefont {Cherstvy}\ \emph {et~al.}(2013)\citenamefont
  {Cherstvy}, \citenamefont {Chechkin},\ and\ \citenamefont
  {Metzler}}]{cherstvy2013}%
  \BibitemOpen
  \bibfield  {author} {\bibinfo {author} {\bibfnamefont {A.~G.}\ \bibnamefont
  {Cherstvy}}, \bibinfo {author} {\bibfnamefont {A.~V.}\ \bibnamefont
  {Chechkin}},\ and\ \bibinfo {author} {\bibfnamefont {R.}~\bibnamefont
  {Metzler}},\ }\bibfield  {title} {\bibinfo {title} {Anomalous diffusion and
  ergodicity breaking in heterogeneous diffusion processes},\ }\href
  {https://doi.org/10.1088/1367-2630/15/8/083039} {\bibfield  {journal}
  {\bibinfo  {journal} {New Journal of Physics}\ }\textbf {\bibinfo {volume}
  {15}},\ \bibinfo {pages} {083039} (\bibinfo {year} {2013})}\BibitemShut
  {NoStop}%
\bibitem [{\citenamefont {Cardy}(2012)}]{cardy2012}%
  \BibitemOpen
  \bibfield  {author} {\bibinfo {author} {\bibfnamefont {J.}~\bibnamefont
  {Cardy}},\ }\href@noop {} {\emph {\bibinfo {title} {Finite-size scaling}}}\
  (\bibinfo  {publisher} {Elsevier},\ \bibinfo {year} {2012})\BibitemShut
  {NoStop}%
\bibitem [{\citenamefont {Kadanoff}(2000)}]{kadanoff2000statistical}%
  \BibitemOpen
  \bibfield  {author} {\bibinfo {author} {\bibfnamefont {L.~P.}\ \bibnamefont
  {Kadanoff}},\ }\href@noop {} {\emph {\bibinfo {title} {Statistical physics:
  statics, dynamics and renormalization}}}\ (\bibinfo  {publisher} {World
  Scientific},\ \bibinfo {year} {2000})\BibitemShut {NoStop}%
\bibitem [{\citenamefont {Touchette}(2009)}]{Touchette2009}%
  \BibitemOpen
  \bibfield  {author} {\bibinfo {author} {\bibfnamefont {H.}~\bibnamefont
  {Touchette}},\ }\bibfield  {title} {\bibinfo {title} {The large deviation
  approach to statistical mechanics},\ }\href
  {https://doi.org/https://doi.org/10.1016/j.physrep.2009.05.002} {\bibfield
  {journal} {\bibinfo  {journal} {Physics Reports}\ }\textbf {\bibinfo {volume}
  {478}},\ \bibinfo {pages} {1} (\bibinfo {year} {2009})}\BibitemShut {NoStop}%
\bibitem [{\citenamefont {Touchette}\ and\ \citenamefont
  {Harris}(2013)}]{touchette2013}%
  \BibitemOpen
  \bibfield  {author} {\bibinfo {author} {\bibfnamefont {H.}~\bibnamefont
  {Touchette}}\ and\ \bibinfo {author} {\bibfnamefont {R.~J.}\ \bibnamefont
  {Harris}},\ }\bibinfo {title} {Large deviation approach to nonequilibrium
  systems},\ in\ \href {https://doi.org/10.1002/9783527658701.ch11} {\emph
  {\bibinfo {booktitle} {Nonequilibrium Statistical Physics of Small
  Systems}}},\ \bibinfo {editor} {edited by\ \bibinfo {editor} {\bibfnamefont
  {R.}~\bibnamefont {Klages}}, \bibinfo {editor} {\bibfnamefont
  {W.}~\bibnamefont {Just}},\ and\ \bibinfo {editor} {\bibfnamefont
  {C.}~\bibnamefont {Jarzynski}}}\ (\bibinfo  {publisher} {John Wiley \& Sons,
  Ltd},\ \bibinfo {year} {2013})\ Chap.~\bibinfo {chapter} {11}, pp.\ \bibinfo
  {pages} {335--360}\BibitemShut {NoStop}%
\bibitem [{\citenamefont {Teza}\ \emph {et~al.}(2022)\citenamefont {Teza},
  \citenamefont {Yaacoby},\ and\ \citenamefont {Raz}}]{teza2022eigenvalue}%
  \BibitemOpen
  \bibfield  {author} {\bibinfo {author} {\bibfnamefont {G.}~\bibnamefont
  {Teza}}, \bibinfo {author} {\bibfnamefont {R.}~\bibnamefont {Yaacoby}},\ and\
  \bibinfo {author} {\bibfnamefont {O.}~\bibnamefont {Raz}},\ }\bibfield
  {title} {\bibinfo {title} {Eigenvalue crossing as a phase transition in
  relaxation dynamics},\ }\bibfield  {journal} {\bibinfo  {journal} {arXiv
  preprint arXiv:2209.09307}\ }\href
  {https://doi.org/10.48550/arXiv.2209.09307} {10.48550/arXiv.2209.09307}
  (\bibinfo {year} {2022})\BibitemShut {NoStop}%
\bibitem [{\citenamefont {Montroll}\ and\ \citenamefont
  {Weiss}(1965)}]{montroll1965random}%
  \BibitemOpen
  \bibfield  {author} {\bibinfo {author} {\bibfnamefont {E.~W.}\ \bibnamefont
  {Montroll}}\ and\ \bibinfo {author} {\bibfnamefont {G.~H.}\ \bibnamefont
  {Weiss}},\ }\bibfield  {title} {\bibinfo {title} {Random walks on lattices.
  ii},\ }\href {https://doi.org/10.1063/1.1704269} {\bibfield  {journal}
  {\bibinfo  {journal} {Journal of Mathematical Physics}\ }\textbf {\bibinfo
  {volume} {6}},\ \bibinfo {pages} {167} (\bibinfo {year} {1965})}\BibitemShut
  {NoStop}%
\bibitem [{\citenamefont {Kenkre}\ \emph {et~al.}(1973)\citenamefont {Kenkre},
  \citenamefont {Montroll},\ and\ \citenamefont
  {Shlesinger}}]{kenkre1973generalized}%
  \BibitemOpen
  \bibfield  {author} {\bibinfo {author} {\bibfnamefont {V.}~\bibnamefont
  {Kenkre}}, \bibinfo {author} {\bibfnamefont {E.}~\bibnamefont {Montroll}},\
  and\ \bibinfo {author} {\bibfnamefont {M.}~\bibnamefont {Shlesinger}},\
  }\bibfield  {title} {\bibinfo {title} {Generalized master equations for
  continuous-time random walks},\ }\href {https://doi.org/10.1007/BF01016796}
  {\bibfield  {journal} {\bibinfo  {journal} {Journal of Statistical Physics}\
  }\textbf {\bibinfo {volume} {9}},\ \bibinfo {pages} {45} (\bibinfo {year}
  {1973})}\BibitemShut {NoStop}%
\bibitem [{\citenamefont {Denisov}\ and\ \citenamefont
  {Horsthemke}(2002{\natexlab{a}})}]{denisov2002statistical}%
  \BibitemOpen
  \bibfield  {author} {\bibinfo {author} {\bibfnamefont {S.~I.}\ \bibnamefont
  {Denisov}}\ and\ \bibinfo {author} {\bibfnamefont {W.}~\bibnamefont
  {Horsthemke}},\ }\bibfield  {title} {\bibinfo {title} {Statistical properties
  of a class of nonlinear systems driven by colored multiplicative gaussian
  noise},\ }\href {https://doi.org/10.1103/PhysRevE.65.031105} {\bibfield
  {journal} {\bibinfo  {journal} {Phys. Rev. E}\ }\textbf {\bibinfo {volume}
  {65}},\ \bibinfo {pages} {031105} (\bibinfo {year}
  {2002}{\natexlab{a}})}\BibitemShut {NoStop}%
\bibitem [{\citenamefont {Denisov}\ and\ \citenamefont
  {Horsthemke}(2002{\natexlab{b}})}]{denisov2002exactly}%
  \BibitemOpen
  \bibfield  {author} {\bibinfo {author} {\bibfnamefont {S.~I.}\ \bibnamefont
  {Denisov}}\ and\ \bibinfo {author} {\bibfnamefont {W.}~\bibnamefont
  {Horsthemke}},\ }\bibfield  {title} {\bibinfo {title} {Exactly solvable model
  with an absorbing state and multiplicative colored gaussian noise},\ }\href
  {https://doi.org/10.1103/PhysRevE.65.061109} {\bibfield  {journal} {\bibinfo
  {journal} {Phys. Rev. E}\ }\textbf {\bibinfo {volume} {65}},\ \bibinfo
  {pages} {061109} (\bibinfo {year} {2002}{\natexlab{b}})}\BibitemShut
  {NoStop}%
\bibitem [{\citenamefont {Sandev}\ \emph
  {et~al.}(2022{\natexlab{a}})\citenamefont {Sandev}, \citenamefont
  {Domazetoski}, \citenamefont {Kocarev}, \citenamefont {Metzler},\ and\
  \citenamefont {Chechkin}}]{sandev2022heterogeneous}%
  \BibitemOpen
  \bibfield  {author} {\bibinfo {author} {\bibfnamefont {T.}~\bibnamefont
  {Sandev}}, \bibinfo {author} {\bibfnamefont {V.}~\bibnamefont {Domazetoski}},
  \bibinfo {author} {\bibfnamefont {L.}~\bibnamefont {Kocarev}}, \bibinfo
  {author} {\bibfnamefont {R.}~\bibnamefont {Metzler}},\ and\ \bibinfo {author}
  {\bibfnamefont {A.}~\bibnamefont {Chechkin}},\ }\bibfield  {title} {\bibinfo
  {title} {Heterogeneous diffusion with stochastic resetting},\ }\href
  {https://doi.org/10.1088/1751-8121/ac491c} {\bibfield  {journal} {\bibinfo
  {journal} {Journal of Physics A: Mathematical and Theoretical}\ }\textbf
  {\bibinfo {volume} {55}},\ \bibinfo {pages} {074003} (\bibinfo {year}
  {2022}{\natexlab{a}})}\BibitemShut {NoStop}%
\bibitem [{\citenamefont {Sandev}\ \emph
  {et~al.}(2022{\natexlab{b}})\citenamefont {Sandev}, \citenamefont {Kocarev},
  \citenamefont {Metzler},\ and\ \citenamefont
  {Chechkin}}]{sandev2022stochastic}%
  \BibitemOpen
  \bibfield  {author} {\bibinfo {author} {\bibfnamefont {T.}~\bibnamefont
  {Sandev}}, \bibinfo {author} {\bibfnamefont {L.}~\bibnamefont {Kocarev}},
  \bibinfo {author} {\bibfnamefont {R.}~\bibnamefont {Metzler}},\ and\ \bibinfo
  {author} {\bibfnamefont {A.}~\bibnamefont {Chechkin}},\ }\bibfield  {title}
  {\bibinfo {title} {Stochastic dynamics with multiplicative dichotomic noise:
  Heterogeneous telegrapher’s equation, anomalous crossovers and resetting},\
  }\href {https://doi.org/https://doi.org/10.1016/j.chaos.2022.112878}
  {\bibfield  {journal} {\bibinfo  {journal} {Chaos, Solitons \& Fractals}\
  }\textbf {\bibinfo {volume} {165}},\ \bibinfo {pages} {112878} (\bibinfo
  {year} {2022}{\natexlab{b}})}\BibitemShut {NoStop}%
\bibitem [{Note1()}]{Note1}%
  \BibitemOpen
  \bibinfo {note} {The case $\nu >1$ correspond to ``hyperballistic''
  diffusion, in which the integral defining the generating function in Eq. \ref
  {eq:gen_func} diverges. We notice how, even though the Richardson class
  encompasses both regimes, the original Richardson model falls in the
  hyperballistic regime with $\nu =3/2$ \cite
  {richardson1926atmospheric,falkovich2001particles,boffetta2002relative}.}\BibitemShut
  {Stop}%
\bibitem [{\citenamefont {Teza}\ and\ \citenamefont
  {Stella}(2020)}]{teza2020exact}%
  \BibitemOpen
  \bibfield  {author} {\bibinfo {author} {\bibfnamefont {G.}~\bibnamefont
  {Teza}}\ and\ \bibinfo {author} {\bibfnamefont {A.~L.}\ \bibnamefont
  {Stella}},\ }\bibfield  {title} {\bibinfo {title} {Exact coarse graining
  preserves entropy production out of equilibrium},\ }\href
  {https://doi.org/10.1103/PhysRevLett.125.110601} {\bibfield  {journal}
  {\bibinfo  {journal} {Phys. Rev. Lett.}\ }\textbf {\bibinfo {volume} {125}},\
  \bibinfo {pages} {110601} (\bibinfo {year} {2020})}\BibitemShut {NoStop}%
\bibitem [{\citenamefont {Teza}(2020)}]{teza2020thesis}%
  \BibitemOpen
  \bibfield  {author} {\bibinfo {author} {\bibfnamefont {G.}~\bibnamefont
  {Teza}},\ }\emph {\bibinfo {title} {Out of equilibrium dynamics: from an
  entropy of the growth to the growth of entropy production}},\ \href
  {http://paduaresearch.cab.unipd.it/12995/} {Ph.D. thesis},\ \bibinfo
  {school} {University of Padova} (\bibinfo {year} {2020})\BibitemShut
  {NoStop}%
\bibitem [{\citenamefont {Bruce}(1995)}]{Bruce1995}%
  \BibitemOpen
  \bibfield  {author} {\bibinfo {author} {\bibfnamefont {A.}~\bibnamefont
  {Bruce}},\ }\bibfield  {title} {\bibinfo {title} {Critical finite-size
  scaling of the free energy},\ }\href
  {https://doi.org/10.1088/0305-4470/28/12/008} {\bibfield  {journal} {\bibinfo
   {journal} {Journal of Physics A: Mathematical and General}\ }\textbf
  {\bibinfo {volume} {28}},\ \bibinfo {pages} {3345} (\bibinfo {year}
  {1995})}\BibitemShut {NoStop}%
\bibitem [{\citenamefont {Privman}\ and\ \citenamefont
  {Fisher}(1984)}]{Privman1984}%
  \BibitemOpen
  \bibfield  {author} {\bibinfo {author} {\bibfnamefont {V.}~\bibnamefont
  {Privman}}\ and\ \bibinfo {author} {\bibfnamefont {M.~E.}\ \bibnamefont
  {Fisher}},\ }\bibfield  {title} {\bibinfo {title} {Universal critical
  amplitudes in finite-size scaling},\ }\href
  {https://doi.org/10.1103/PhysRevB.30.322} {\bibfield  {journal} {\bibinfo
  {journal} {Phys. Rev. B}\ }\textbf {\bibinfo {volume} {30}},\ \bibinfo
  {pages} {322} (\bibinfo {year} {1984})}\BibitemShut {NoStop}%
\bibitem [{\citenamefont {Bl{\"o}te}\ \emph {et~al.}(1986)\citenamefont
  {Bl{\"o}te}, \citenamefont {Cardy},\ and\ \citenamefont
  {Nightingale}}]{blote1986}%
  \BibitemOpen
  \bibfield  {author} {\bibinfo {author} {\bibfnamefont {H.~W.}\ \bibnamefont
  {Bl{\"o}te}}, \bibinfo {author} {\bibfnamefont {J.~L.}\ \bibnamefont
  {Cardy}},\ and\ \bibinfo {author} {\bibfnamefont {M.~P.}\ \bibnamefont
  {Nightingale}},\ }\bibfield  {title} {\bibinfo {title} {Conformal invariance,
  the central charge, and universal finite-size amplitudes at criticality},\
  }\href@noop {} {\bibfield  {journal} {\bibinfo  {journal} {Physical review
  letters}\ }\textbf {\bibinfo {volume} {56}},\ \bibinfo {pages} {742}
  (\bibinfo {year} {1986})}\BibitemShut {NoStop}%
\bibitem [{\citenamefont {G\"{a}rtner}(1977)}]{gartner1977on}%
  \BibitemOpen
  \bibfield  {author} {\bibinfo {author} {\bibfnamefont {J.}~\bibnamefont
  {G\"{a}rtner}},\ }\bibfield  {title} {\bibinfo {title} {On large deviations
  from the invariant measure},\ }\href {https://doi.org/10.1137/1122003}
  {\bibfield  {journal} {\bibinfo  {journal} {Theory of Probability \& Its
  Applications}\ }\textbf {\bibinfo {volume} {22}},\ \bibinfo {pages} {24}
  (\bibinfo {year} {1977})},\ \Eprint
  {https://arxiv.org/abs/https://doi.org/10.1137/1122003}
  {https://doi.org/10.1137/1122003} \BibitemShut {NoStop}%
\bibitem [{\citenamefont {Ellis}(1984)}]{ellis1984large}%
  \BibitemOpen
  \bibfield  {author} {\bibinfo {author} {\bibfnamefont {R.~S.}\ \bibnamefont
  {Ellis}},\ }\bibfield  {title} {\bibinfo {title} {Large deviations for a
  general class of random vectors},\ }\href
  {http://www.jstor.org/stable/2243592} {\bibfield  {journal} {\bibinfo
  {journal} {The Annals of Probability}\ }\textbf {\bibinfo {volume} {12}},\
  \bibinfo {pages} {1} (\bibinfo {year} {1984})}\BibitemShut {NoStop}%
\bibitem [{\citenamefont {Rockafellar}(1970)}]{rockafellar1970convex}%
  \BibitemOpen
  \bibfield  {author} {\bibinfo {author} {\bibfnamefont {R.~T.}\ \bibnamefont
  {Rockafellar}},\ }\href@noop {} {\emph {\bibinfo {title} {Convex
  analysis}}},\ Vol.~\bibinfo {volume} {18}\ (\bibinfo  {publisher} {Princeton
  university press},\ \bibinfo {year} {1970})\BibitemShut {NoStop}%
\bibitem [{\citenamefont {Teza}\ \emph {et~al.}(2020)\citenamefont {Teza},
  \citenamefont {Iubini}, \citenamefont {Baiesi}, \citenamefont {Stella},\ and\
  \citenamefont {Vanderzande}}]{teza2020rate}%
  \BibitemOpen
  \bibfield  {author} {\bibinfo {author} {\bibfnamefont {G.}~\bibnamefont
  {Teza}}, \bibinfo {author} {\bibfnamefont {S.}~\bibnamefont {Iubini}},
  \bibinfo {author} {\bibfnamefont {M.}~\bibnamefont {Baiesi}}, \bibinfo
  {author} {\bibfnamefont {A.~L.}\ \bibnamefont {Stella}},\ and\ \bibinfo
  {author} {\bibfnamefont {C.}~\bibnamefont {Vanderzande}},\ }\bibfield
  {title} {\bibinfo {title} {Rate dependence of current and fluctuations in
  jump models with negative differential mobility},\ }\href
  {https://doi.org/https://doi.org/10.1016/j.physa.2019.123176} {\bibfield
  {journal} {\bibinfo  {journal} {Physica A: Statistical Mechanics and its
  Applications}\ }\textbf {\bibinfo {volume} {552}},\ \bibinfo {pages} {123176}
  (\bibinfo {year} {2020})},\ \bibinfo {note} {tributes of Non-equilibrium
  Statistical Physics}\BibitemShut {NoStop}%
\bibitem [{\citenamefont {Nickelsen}\ and\ \citenamefont
  {Touchette}(2018)}]{nickelsen2018anomalous}%
  \BibitemOpen
  \bibfield  {author} {\bibinfo {author} {\bibfnamefont {D.}~\bibnamefont
  {Nickelsen}}\ and\ \bibinfo {author} {\bibfnamefont {H.}~\bibnamefont
  {Touchette}},\ }\bibfield  {title} {\bibinfo {title} {Anomalous scaling of
  dynamical large deviations},\ }\href
  {https://doi.org/10.1103/PhysRevLett.121.090602} {\bibfield  {journal}
  {\bibinfo  {journal} {Phys. Rev. Lett.}\ }\textbf {\bibinfo {volume} {121}},\
  \bibinfo {pages} {090602} (\bibinfo {year} {2018})}\BibitemShut {NoStop}%
\bibitem [{\citenamefont {Smith}(2022)}]{smith2022anomalous}%
  \BibitemOpen
  \bibfield  {author} {\bibinfo {author} {\bibfnamefont {N.~R.}\ \bibnamefont
  {Smith}},\ }\bibfield  {title} {\bibinfo {title} {Anomalous scaling and
  first-order dynamical phase transition in large deviations of the
  ornstein-uhlenbeck process},\ }\href
  {https://doi.org/10.1103/PhysRevE.105.014120} {\bibfield  {journal} {\bibinfo
   {journal} {Phys. Rev. E}\ }\textbf {\bibinfo {volume} {105}},\ \bibinfo
  {pages} {014120} (\bibinfo {year} {2022})}\BibitemShut {NoStop}%
\bibitem [{\citenamefont {Nickelsen}\ and\ \citenamefont
  {Touchette}(2022)}]{nickelsen2022noise}%
  \BibitemOpen
  \bibfield  {author} {\bibinfo {author} {\bibfnamefont {D.}~\bibnamefont
  {Nickelsen}}\ and\ \bibinfo {author} {\bibfnamefont {H.}~\bibnamefont
  {Touchette}},\ }\bibfield  {title} {\bibinfo {title} {Noise correction of
  large deviations with anomalous scaling},\ }\href
  {https://doi.org/10.1103/PhysRevE.105.064102} {\bibfield  {journal} {\bibinfo
   {journal} {Phys. Rev. E}\ }\textbf {\bibinfo {volume} {105}},\ \bibinfo
  {pages} {064102} (\bibinfo {year} {2022})}\BibitemShut {NoStop}%
\end{thebibliography}%

\end{document}